\documentclass{main}
\usepackage{textcomp}
\usepackage{fancyhdr}
\usepackage{lastpage}
\def\be{\begin{equation}}
\def\ee{\end{equation}}
\def\bea{\begin{eqnarray}}
\def\eea{\end{eqnarray}}

\newcommand{\Photo}{\includegraphics[height=35mm]{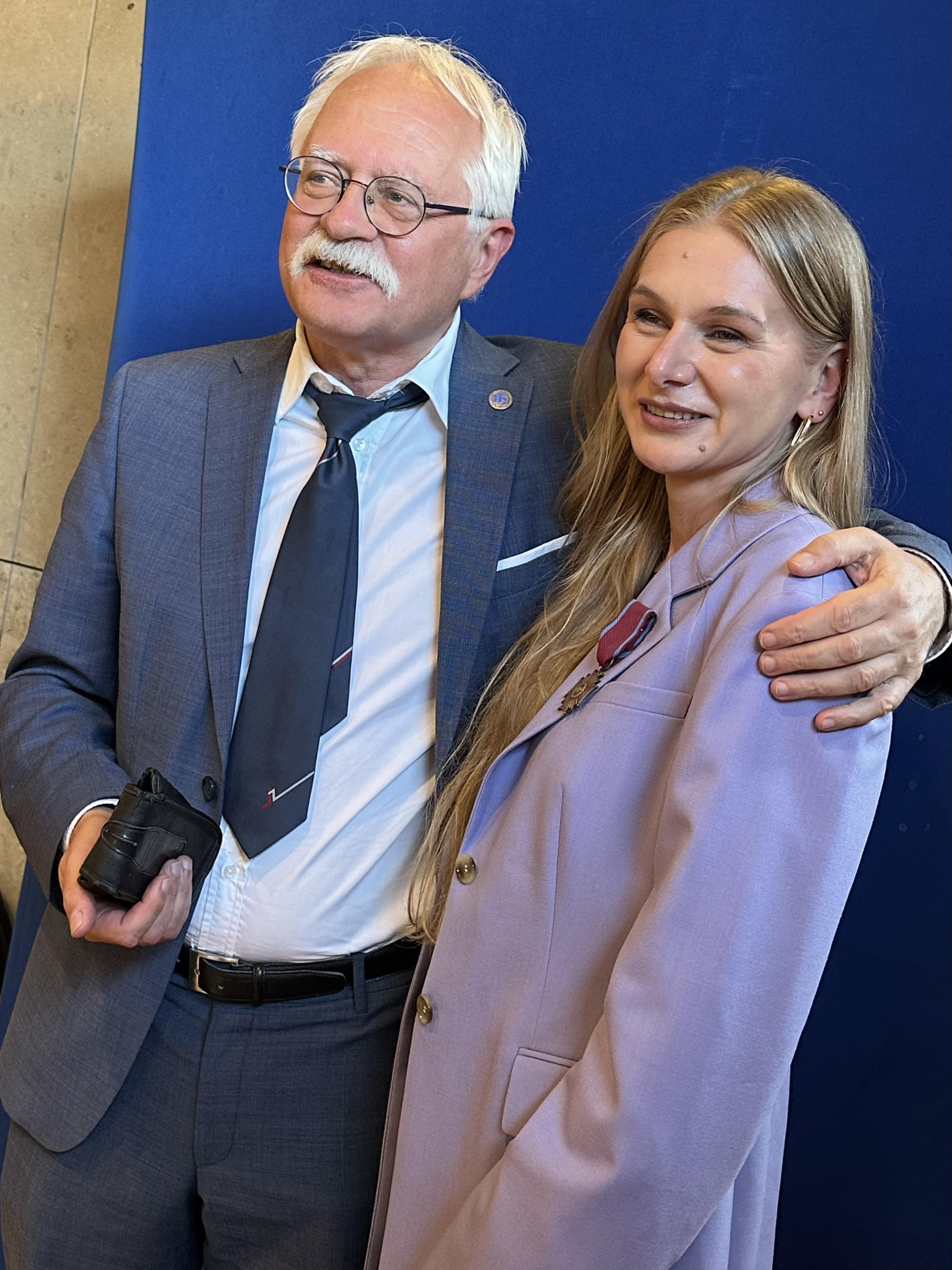}}

\fancypagestyle{firstpagefooter}{%
  \fancyfoot[L]{  \textcopyright \footnotesize This work is an open access article under a Creative Commons Attribution 4.0 International License (https://creativecommons.org/licenses/by/4.0/).}
  \fancyfoot[C]{}  % **Red: This removes the page number from the first page footer**
   
}

\begin{document}

\thispagestyle{firstpagefooter}
\title{\Large Impact of inelastic corrections on 
$\gamma \gamma \to \gamma \gamma$ scattering from UPC at the LHC}

\author{\underline{A. Szczurek}\footnote{Speaker, email: 
antoni.szczurek@ifj.edu.pl}
and M. K{\l}usek-Gawenda}

\address{
Institute of Nuclear Physics, Krakow 31-034, Poland,\\
Institute of Physics, Faculty of Exact and Technical Sciences, University of Rzeszów,
Rzeszów, 35-310, Poland
}

\maketitle\abstracts{
The current theoretical estimations lead to cross-sections
for $AA \to \gamma \gamma AA$ which are somewhat
smaller than the measured ones by the ATLAS and CMS Collaborations,
In our recent paper, we estimated the contribution of inelastic 
channels to the Light-by-Light (LbL) scattering in ultraperipheral 
collisions (UPC) of heavy ions, in which one or both of the 
incident nuclei dissociate  ($A A \to \gamma \gamma X Y$ 
where $X, Y = A, A'$) due to the photon emission. 
These new mechanisms are related to extra emissions that 
are difficult to identify 
and  may be misinterpreted as enhanced 
$\gamma \gamma \to \gamma \gamma$ scattering.
We include processes of coupling of photons to individual nucleons
in addition to coherent coupling to the whole nuclei. 
Both elastic (nucleon in the ground state) and inelastic 
(nucleon in an excited state) are taken into account. 
The inelastic nucleon fluxes are calculated
using CTEQ18QED photon distribution in nucleon. 
The inelastic photon fluxes are shown and compared to 
standard photon fluxes in the nucleus.
We present the ratio of the inelastic corrections to 
the standard contribution to the nuclear cross section.
We find that for the ATLAS kinematics the inelastic 
corrections grow with $M_{\gamma \gamma}$ and rapidity difference. 
Our results indicate that the inelastic contributions can 
be of the order of 20-40 \%  
of the traditional (no nuclear excitation) predictions. 
We discuss also uncertainties due to the choice of 
factorization scale.
}

\footnotesize DOI: \url{https://doi.org/xx.yyyyy/nnnnnnnn}

\keywords{photon-photon scattering, ultraperipheral heavy ion 
collisions, photon flux in nuclei and nucleons}

\newpage

%----------------------------
\section{Introduction}
%----------------------------

\begin{figure}[!h]
\begin{center}
	\includegraphics[scale=0.25]{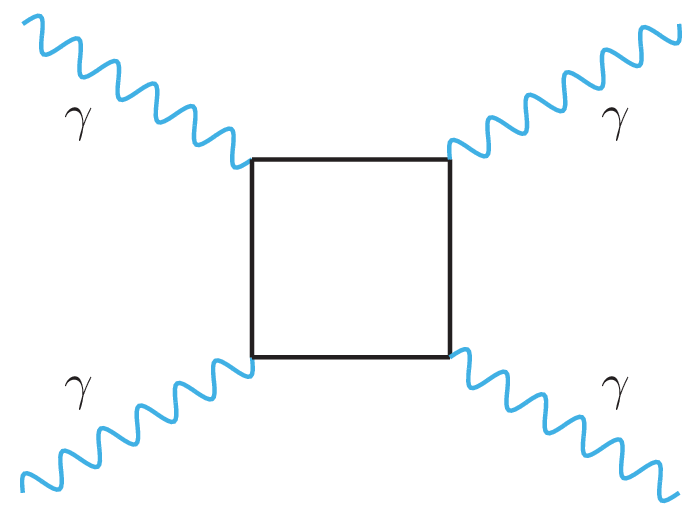}
        \includegraphics[scale=0.25]{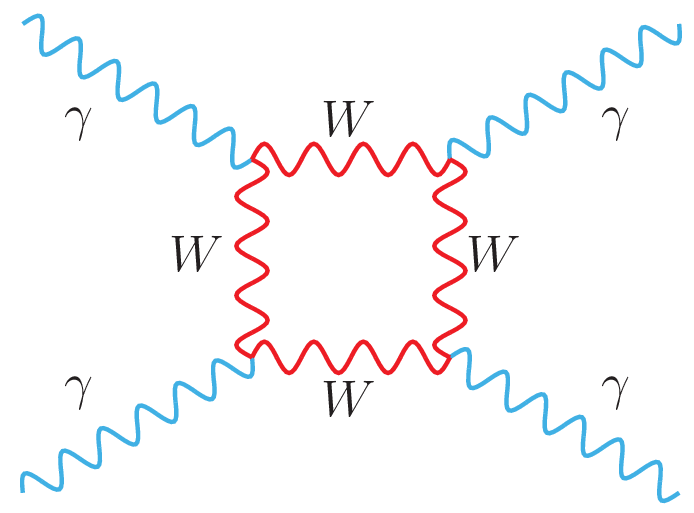} \\
        \includegraphics[scale=0.25]{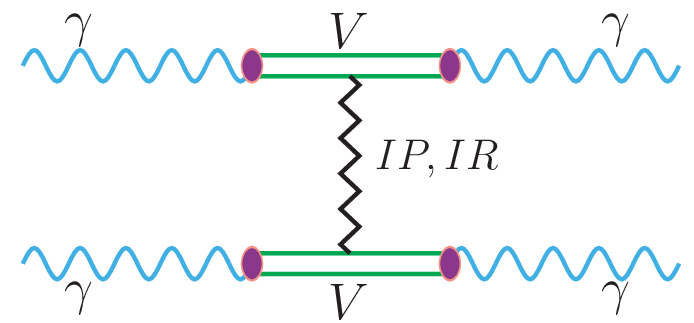}
        \includegraphics[scale=0.25]{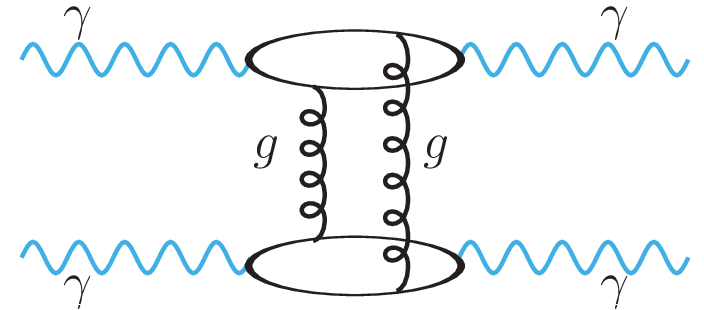}
\end{center}
\caption{Different mechanisms of $\gamma \gamma \to \gamma \gamma$
scattering.}
\label{fig:mechanisms}
\end{figure}

The $\gamma \gamma \to \gamma \gamma$ scattering is fully quantal 
effect. Several mechanisms can be present in principle.
Some of them are shown in Fig. \ref{fig:mechanisms}.
It was/is a dream to observe photon-photon scattering 
for the laser community. Some works in this direction are going on.
An alternative was proposed by D. d'Enterria and G. Silveira to do 
it in UPC of heavy ions \cite{ES2013}.
Only box diagrams were consider in this early evaluation.
M. Klusek-Gawenda, P. Lebiedowicz and A. Szczurek proposed an updade
\cite{Klusek-Gawenda:2016euz}. 
Different mechansims were considered,
including box diagrams as well as
double-hadronic excitations with virtual vector mesons, 
called by us VDM Regge.
Our group considered also two-gluon exchange mechanism in
\cite{KSS2016}.
Both ATLAS and CMS collaborations performed relevant experiments 
and got results roughly consistent with our predictions.

At lower energies the photon-photon scattering may be
contaminated by the background contribution:
$\gamma \gamma \to \pi^0 (\to 2 \gamma) \pi^0 (\to 2 \gamma)$
when only two photons out of four are measured in the fiducial
region. How to eliminate such a background was discussed
in detail e.g. in \cite{Klusek-Gawenda:2019ijn}.
There are at least two experimental projects that could provide new 
insights into photon-photon scattering: FoCal and ALICE 3.
This was discussed in \cite{Jucha:2023hjg}.

%---------------------------------------------------------------------
%---------------------------------------------------------------------

\begin{figure}[!]
       \begin{center}
		\includegraphics[scale=0.3]{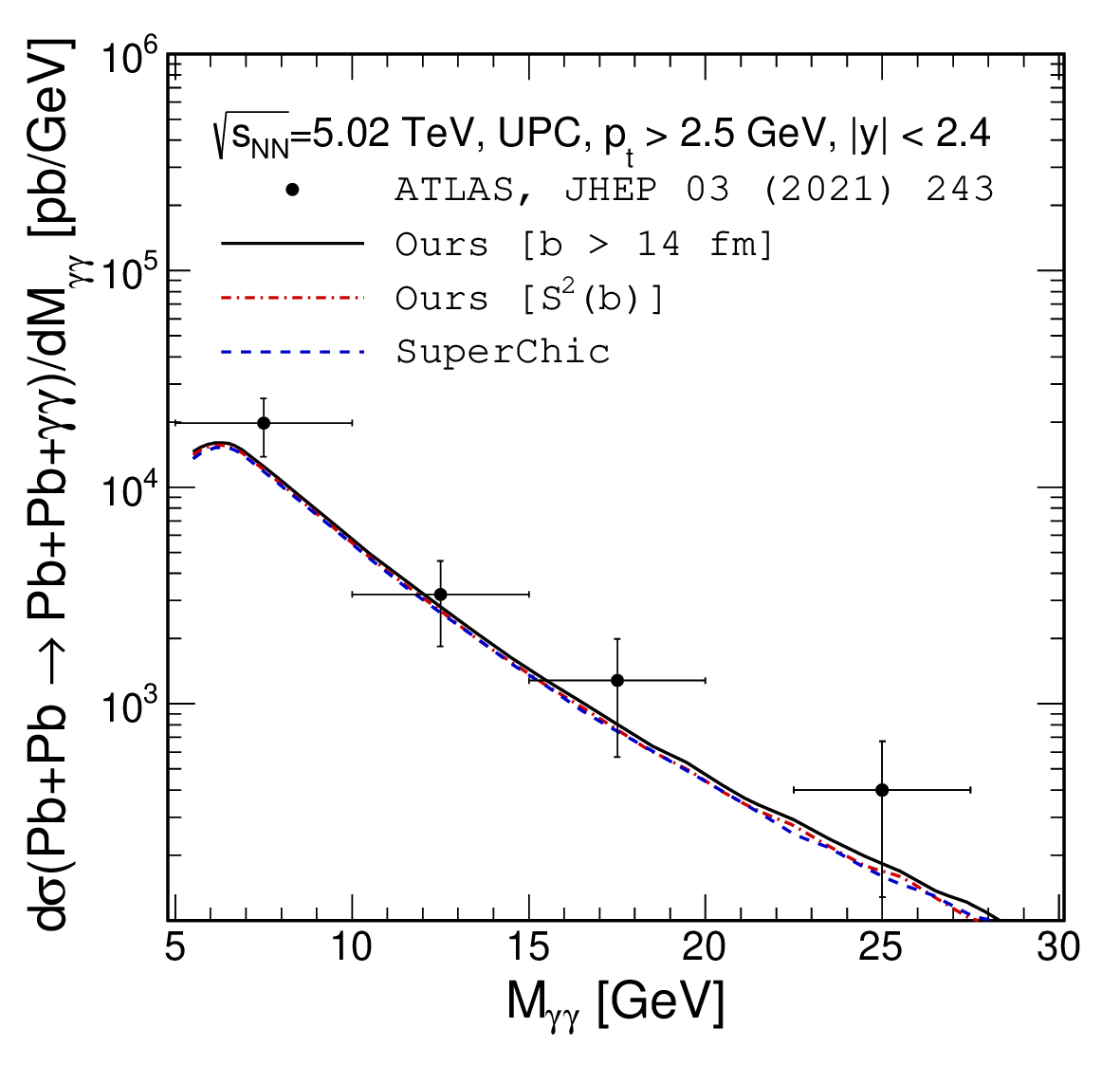}
       \end{center}
\caption{Diphoton invariant mass distribution obtained in different
technical ways. Here only double coherent distribution is included.}
\label{fig:ATLAS}
\end{figure}

The current situation is discussed in Fig. \ref{fig:ATLAS}.
There seem to be some missing strenght. This is even more evident
when discussing integrated cross section, see \cite{Jucha:2023hjg}.
What can be the missing mechanism?

%----------------------------------------------------------------------
%-----------------------------------------------------------------------

%--------------------------------------------
\section{Sketch of the formalism}
%--------------------------------------------

In the state-of-art calculation the nuclear cross section is 
calculated using equivalent photon approximation (EPA) in the b-space. 
In this approach, the di-photon cross section can be written as:
\begin{eqnarray}
    \frac{d\sigma(PbPb \to PbPb \gamma \gamma)}{dy_{\gamma_1}dy_{\gamma_2}dp_{t,\gamma}} &=& 
    \int
\frac{d\sigma_{\gamma\gamma\to\gamma\gamma}(W_{\gamma\gamma})}{dz}
N(\omega_1,b_1) N(\omega_2,b_2) S^2_{abs}(b) \nonumber \\  
    &\times& d^2b d\bar{b}_x d\bar{b}_y \frac{W_{\gamma\gamma}}{2} \frac{dW_{\gamma\gamma}dY_{\gamma\gamma}}{dy_{\gamma_1}dy_{\gamma_2}dp_{t,\gamma}} dz \;,
    \label{eq:tot_xsec}
\end{eqnarray}
where $\bar{b}_x = \left( b_{1x} + b_{2x}\right)/2$ and $\bar{b}_y =
\left( b_{1y} + b_{2y}\right)/2$. The relation between $\vec{b}_1$,
$\vec{b}_2$ and the impact parameter reads: 
$b = |\vec{b}| = \sqrt{|\vec{b}_1|^2 + |\vec{b}_2|^2 - 2|\vec{b}_1||\vec{b}_2|\cos\phi}$. Absorption factor $S^2_{abs}(b)$ is calculated as:

\begin{eqnarray}
S^2_{abs}(b) = \Theta(b-b_{max})   
\end{eqnarray}

\hspace{-0.5cm} or

\begin{eqnarray}
S^2_{abs}(b) = exp\left( -\sigma_{NN} T_{AA}(b) \right)  \;,
    \label{eq:s2b}
\end{eqnarray}

\hspace{-0.5cm} where $\sigma_{NN}$ is the energy-dependent 
nucleon-nucleon interaction cross section.

%--------------------------------------
\section{Inelastic processes}
%--------------------------------------

\begin{figure}[!h]
\begin{center}
   \includegraphics[scale=0.2]{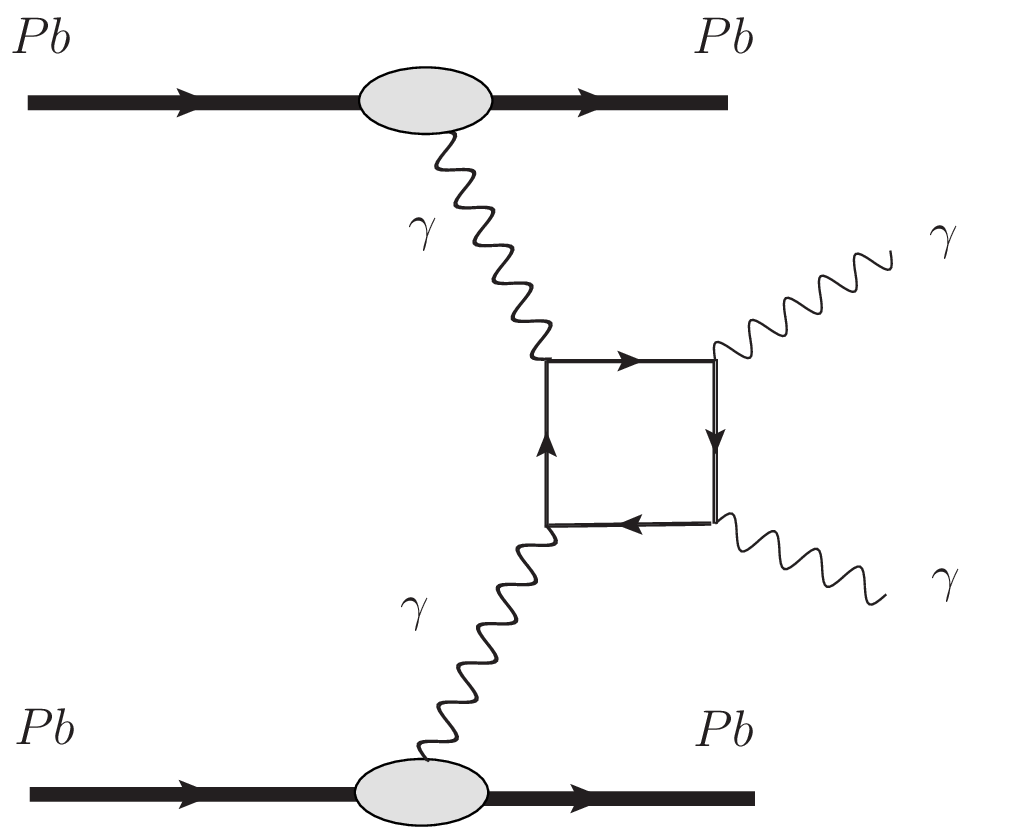}
   \includegraphics[scale=0.2]{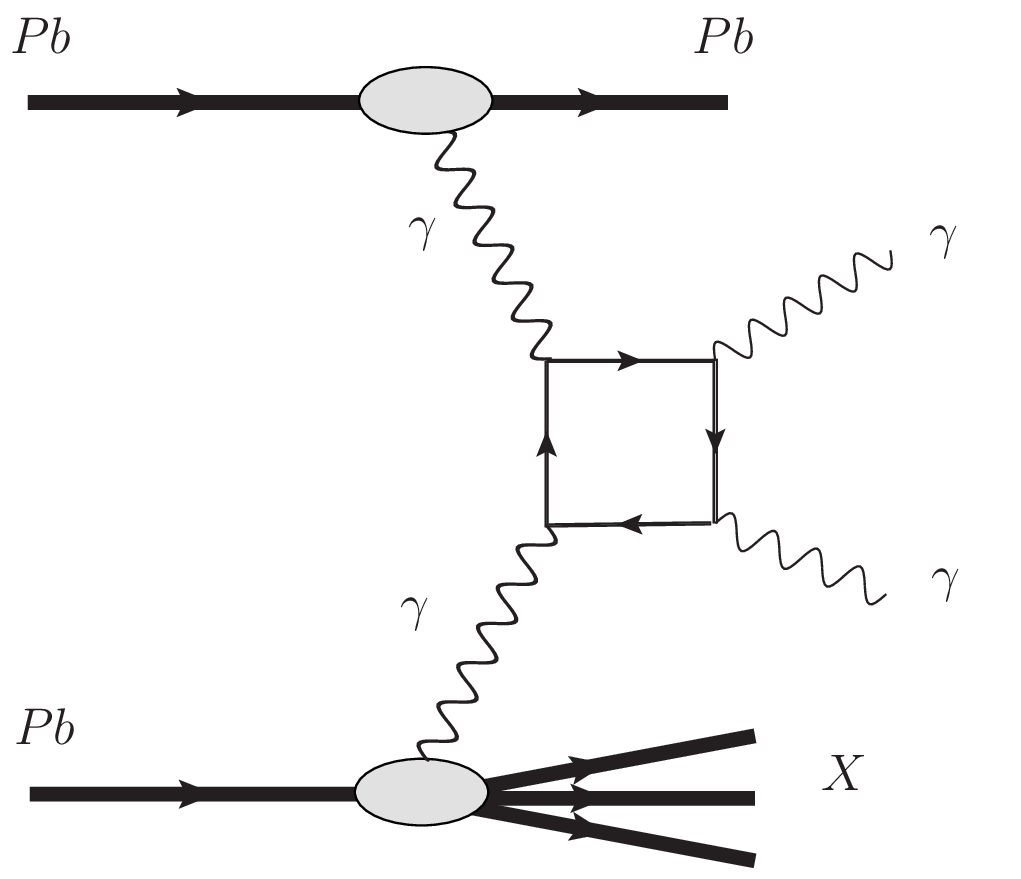}
   \includegraphics[scale=0.2]{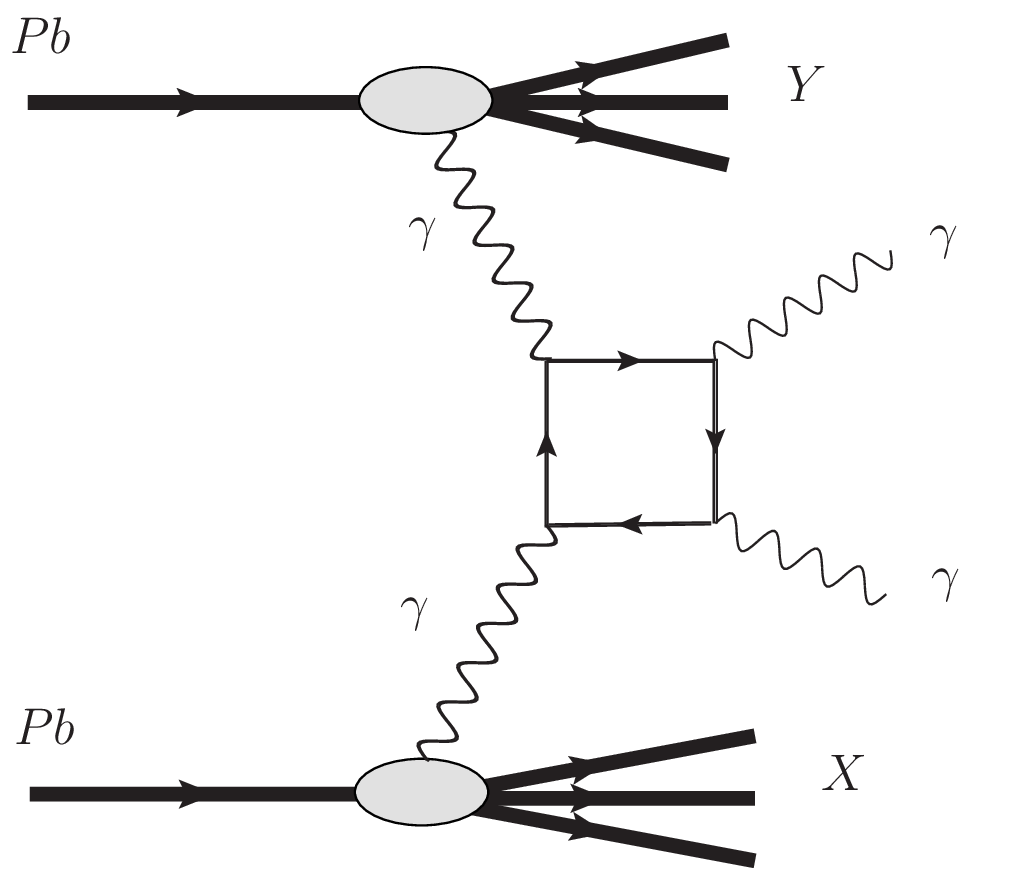}   
\end{center}
\caption{New inelastic mechanisms of diphoton production.}
\label{fig:inelastic}
\end{figure}

In Fig. \ref{fig:inelastic} we show new inelastic mechanisms
included recently in \cite{GKS2025}.

%---------------------------------------------------------------------
%---------------------------------------------------------------------

The inclusion of the inelastic component in the calculation of LbL 
scattering in $PbPb$ collisions implies that the cross-section 
will be given schematically by 
\begin{eqnarray}
\sigma^{LbL} (\sqrt{s_{NN}})   
&\propto&    f_{\gamma/Pb}^{el}(x_1) \otimes f_{\gamma/Pb}^{el}(x_2)  \otimes  \, \hat{\sigma}\left[\gamma \gamma \rightarrow \gamma \gamma ; W_{\gamma \gamma} \right]  \,\,  \nonumber \\
& + & f_{\gamma/Pb}^{el}(x_1) \otimes f_{\gamma/Pb}^{inel}(x_2)  \otimes  \, \hat{\sigma}\left[\gamma \gamma \rightarrow \gamma \gamma ; W_{\gamma \gamma} \right]  \,\,  \nonumber \\
& + & f_{\gamma/Pb}^{inel}(x_1) \otimes f_{\gamma/Pb}^{el}(x_2)  \otimes  \, \hat{\sigma}\left[\gamma \gamma \rightarrow \gamma \gamma ; W_{\gamma \gamma} \right]  \,\,  \nonumber \\
& + & f_{\gamma/Pb}^{inel}(x_1) \otimes f_{\gamma/Pb}^{inel}(x_2)  \otimes  \, \hat{\sigma}\left[\gamma \gamma \rightarrow \gamma \gamma ; W_{\gamma \gamma} \right]   \,\,\, ,
\label{Eq:LbL}
\end{eqnarray}
In the simlified calculation in \cite{GKS2025} the fluxes of photons 
are calculated as:
\begin{eqnarray}
f_{\gamma/Pb}^{el} (x) & = & \frac{\alpha Z^2}{\pi x}\left\{2\xi K_0(\xi)K_1(\xi) - \xi^2[K_1^2(\xi) - K_0^2(\xi)] \right\}  \,\,,  
\end{eqnarray}
and
\begin{eqnarray}
f_{\gamma/Pb}^{inel} (x,\mu^2) & = & 
Z \times f_{\gamma/p} (x,\mu^2) + 
(A - Z) \times f_{\gamma/n} (x,\mu^2) \,\,,
\end{eqnarray}

Photon distribution in nucleon is obtained by solving 
corrected DGLAP equation.
In \cite{GKS2025} both photon in proton 
and photon in neutron were taken from CTEQ18QED
\cite{Xie:2021equ,Xie:2023qbn}.

\begin{eqnarray}
f_{\gamma/p}(x,\mu^2) &=& f_{\gamma/p}^{el}(x) +
f_{\gamma/p}^{inel}(x,\mu^2) , \nonumber \\
f_{\gamma/n}(x,\mu^2) &=& f_{\gamma/n}^{el}(x) + 
f_{\gamma/n}^{inel}(x,\mu^2) . \nonumber 
\end{eqnarray}
%

%--------------------------------------------------------
\begin{figure}[!h]
\begin{center}
    \includegraphics[scale=0.3]{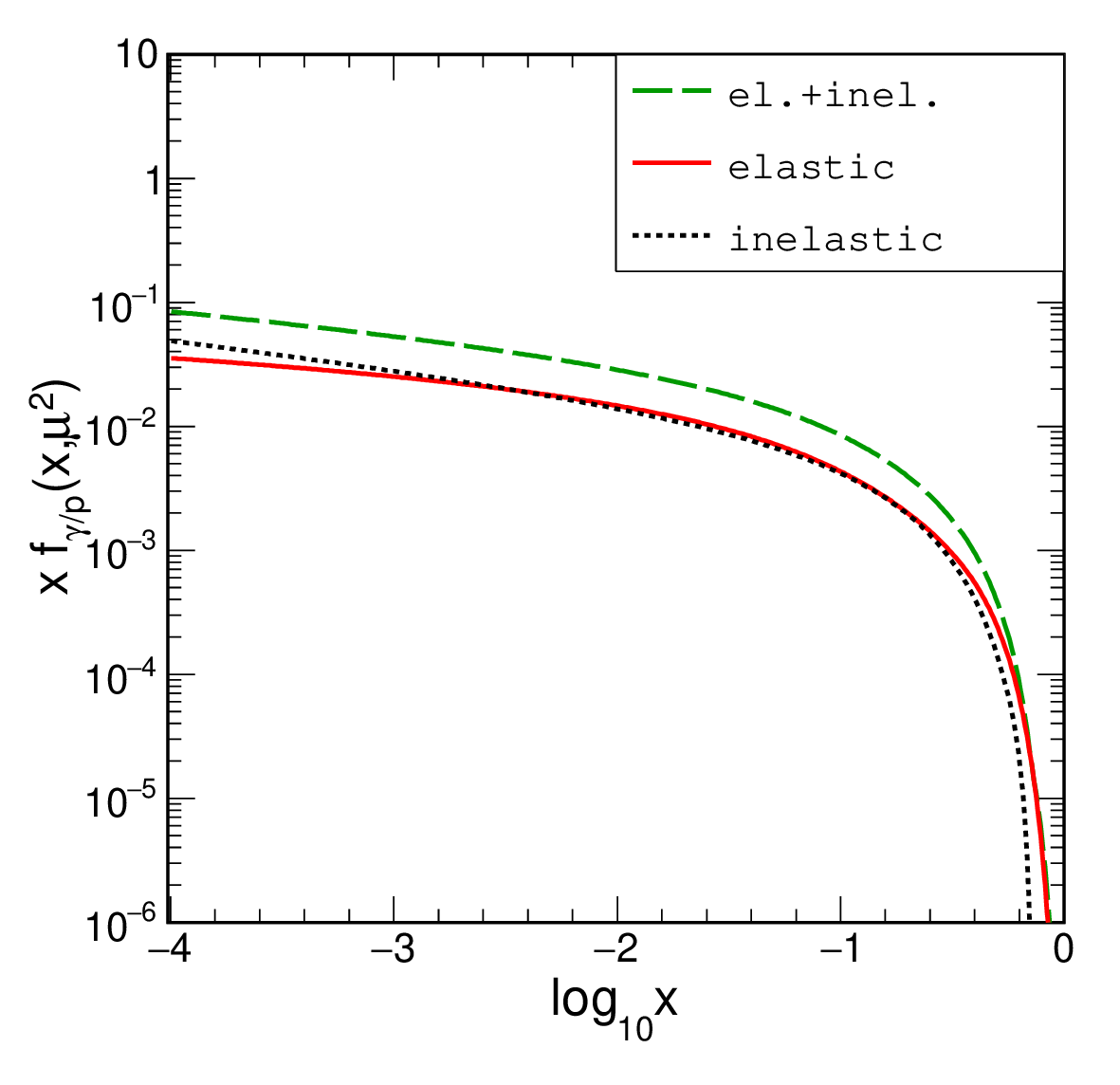}
    \includegraphics[scale=0.3]{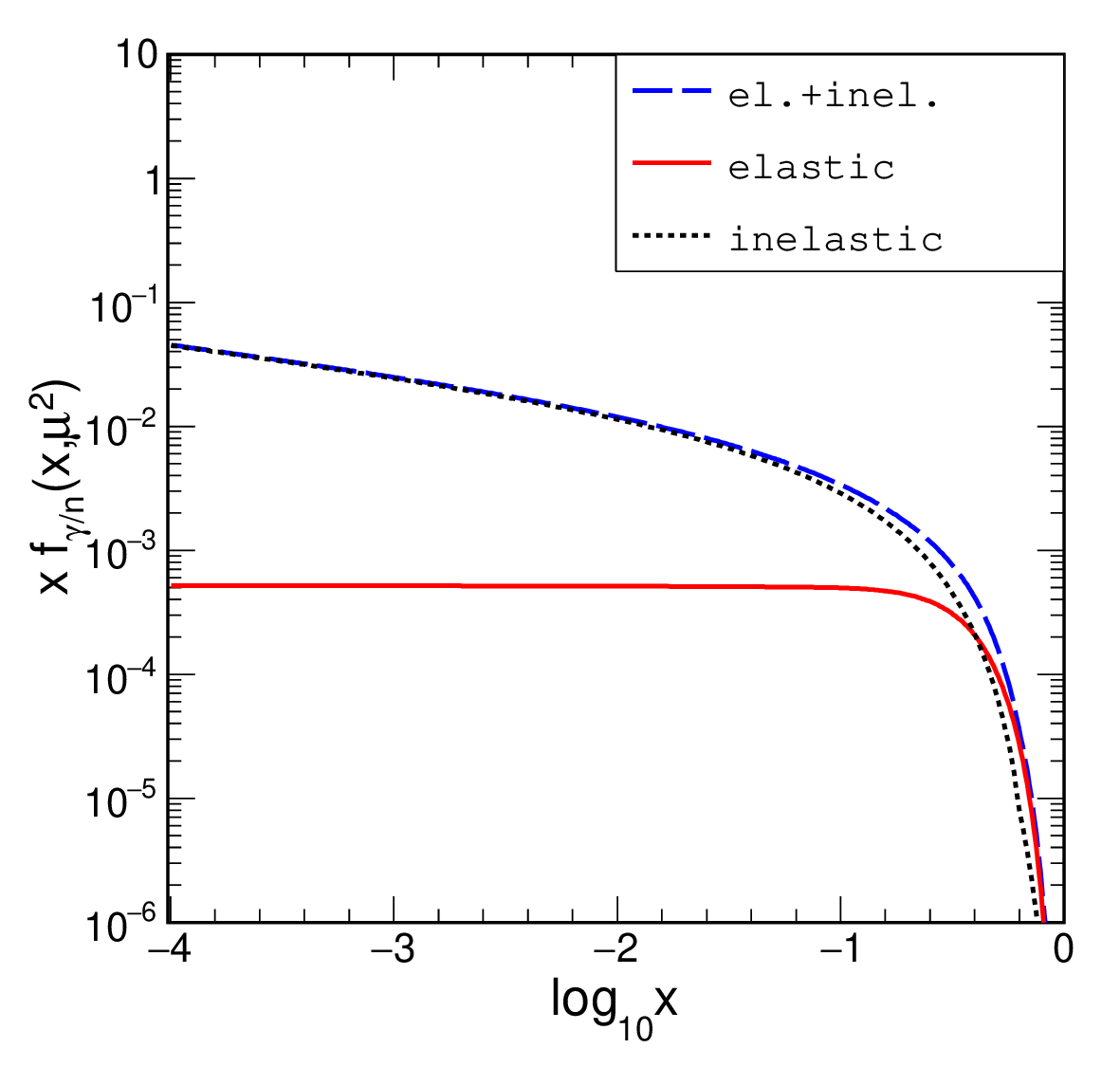}
\end{center}
\caption{$x f_{\gamma/p}(x,\mu^2)$ (left) and
         $x f_{\gamma/n}(x,\mu^2)$ (right).
Here $\mu$ = 5 GeV.}
\label{fig:xgamma_logx}
\end{figure}
%--------------------------------------------------------

The photon distributions as a function of $\log_{10}(x)$
for fixed factorization scale are shown in Fig. \ref{fig:xgamma_logx}.
There are two components in $f_{\gamma/p}$ and $f_{\gamma/n}$.
Here we use CTEQ18QED photon fluxes \cite{{Xie:2021equ},Xie:2023qbn}.
The elastic part is described in terms of electric and magnetic form factors; for neutrons (right panel) the magnetic form factor dominates due to their zero electric charge, whereas for protons (left panel) both form factors play significant role. For protons, the elastic and inelastic contributions are of comparable size, while for neutrons the elastic part is very small.

%--------------------------------------------------------
\begin{figure}[!h]
\begin{center}
    \includegraphics[scale=0.3]{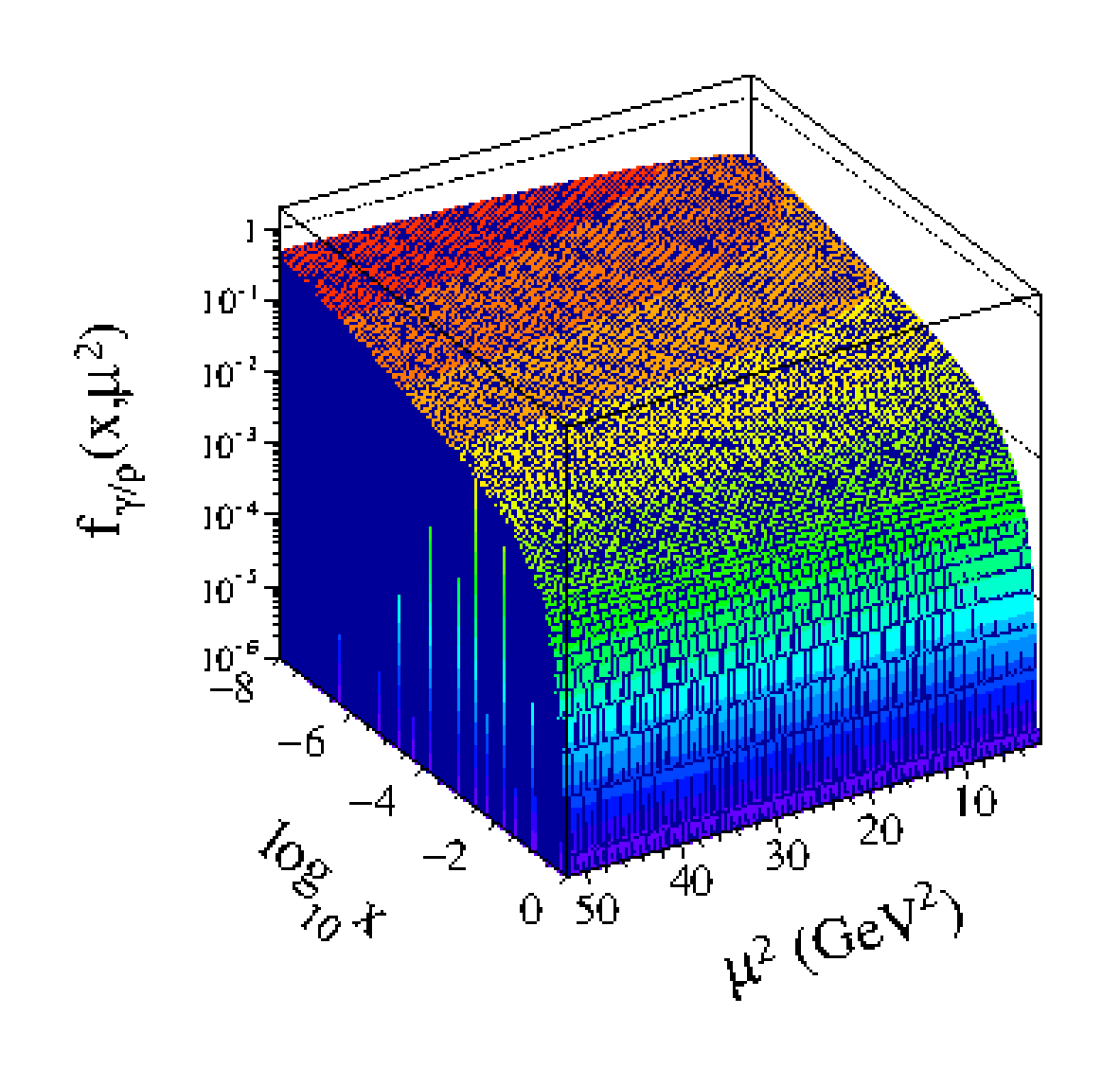}
    \includegraphics[scale=0.3]{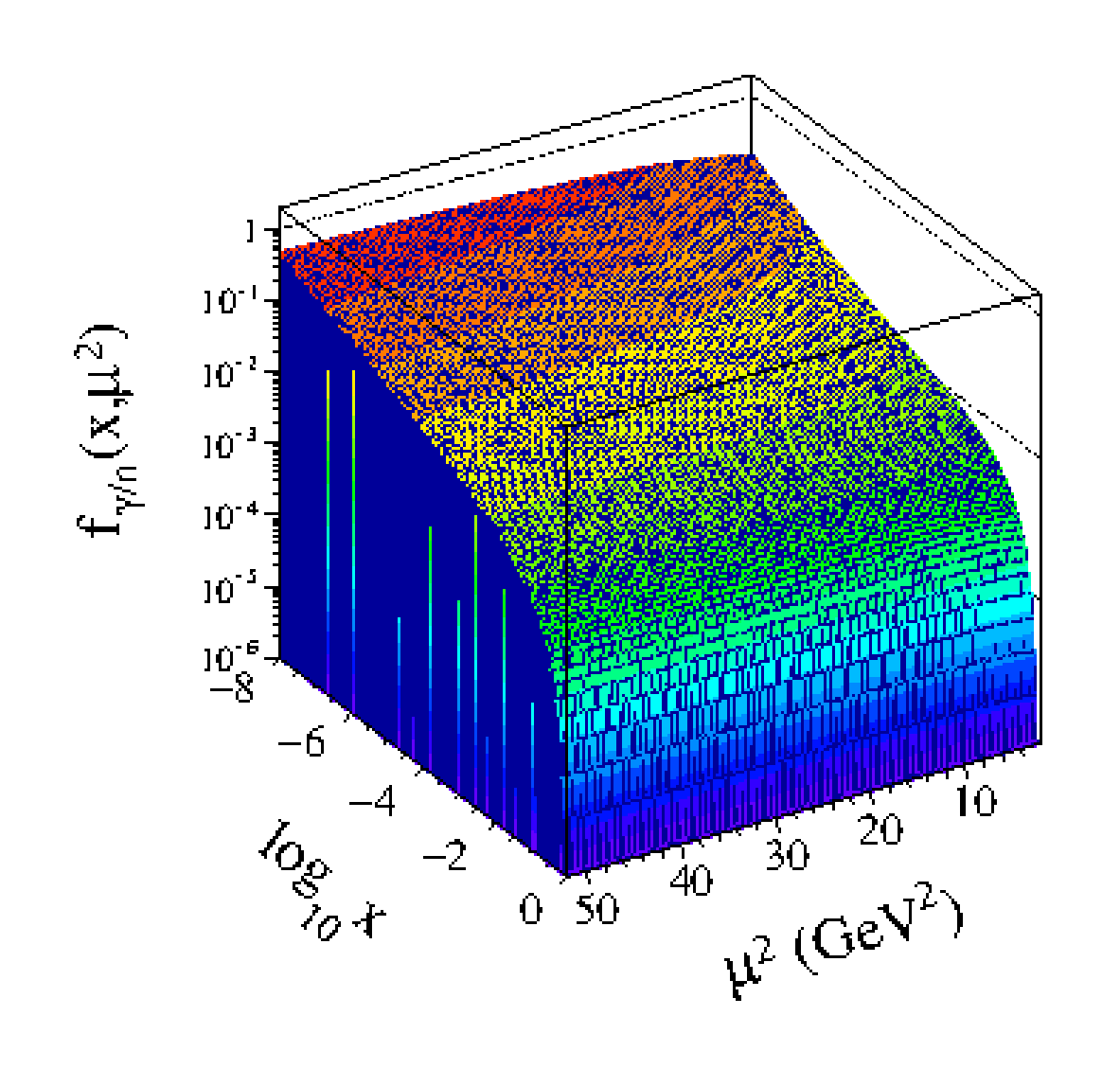}
\end{center}
\caption{$x f_{\gamma/p}(x,\mu^2)$ (left) and
         $x f_{\gamma/n}(x,\mu^2)$ (right).}
\label{fig:f_gamma_ximu2}
\end{figure}
%--------------------------------------------------------

%There are two components in $f_{\gamma/p}$ and $f_{\gamma/n}$.
%In the recent analysis we used CTEQ18QED photon fluxes.

In Fig. \ref{fig:f_gamma_ximu2}, the photon distributions in the proton (left panel) and neutron (right panel) are shown as functions of $\log_{10}x$ and the $\mu^2$ scale. It can be observed that the dependence on the scale is relatively light. The most pronounced differences between the proton and neutron photon PDFs arise at low hard scales, which can be attributed to the distinct elastic photon flux contributions, and at large values of $x$, where the effect is driven by the differences in the quark content of the two nucleons.

%------------------------------------------------------
\begin{figure}[!h]
   \begin{center}
     \includegraphics[scale=0.3]{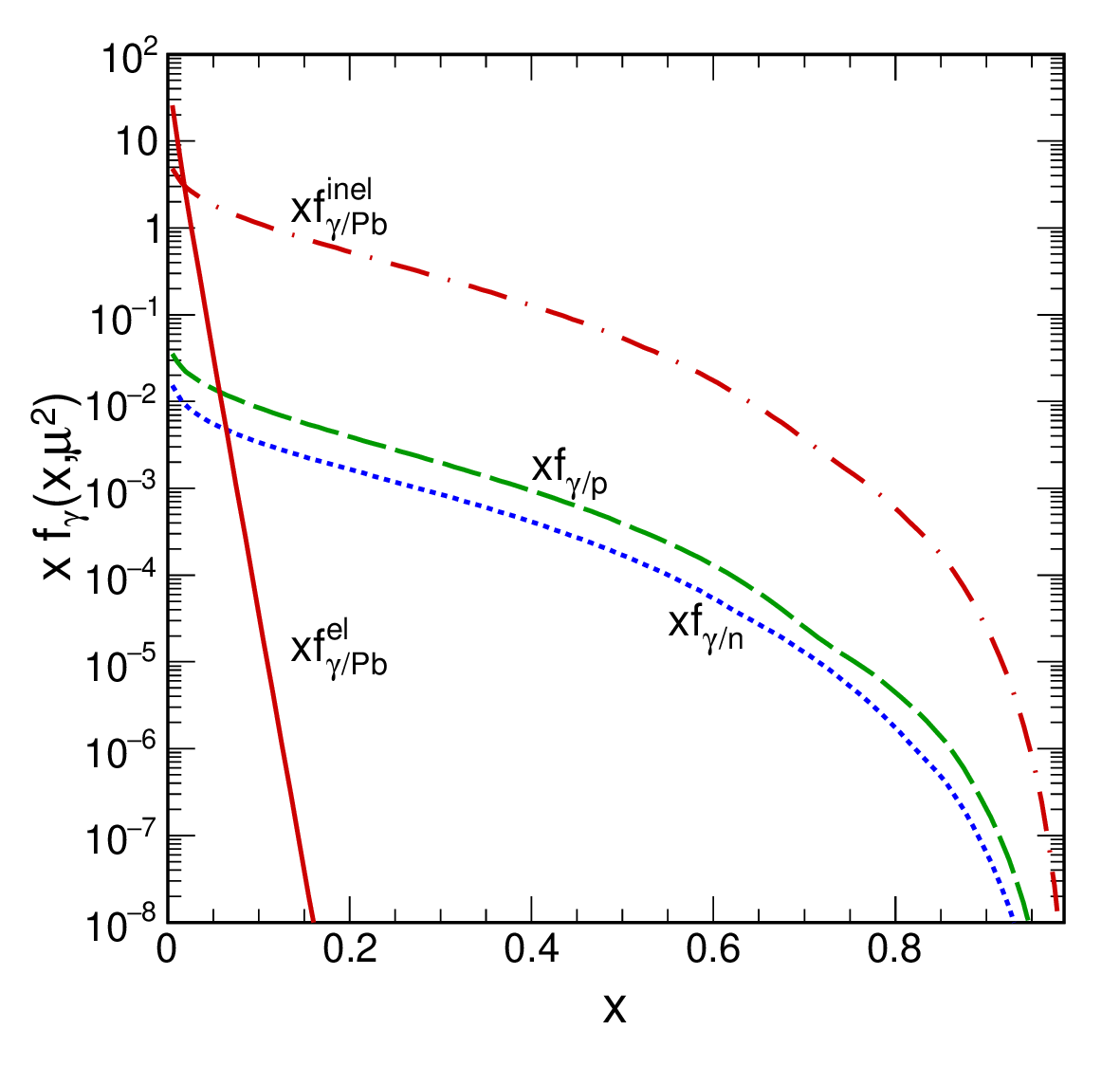}
   \end{center}
\caption{$x f_{\gamma/Pb}(x,\mu=5\, GeV)$ elastic and inelastic 
contributions.}
\label{fig:xfgamma_x_in_A}
\end{figure}
%------------------------------------------------------

The elastic (standard) and inelatic (new) photon distributions in 
$^{208}Pb$, along with proton and neutron fluxes for reference 
are shown separately in Fig.~\ref{fig:xfgamma_x_in_A}. 
The inelastic flux scales with $A$, while the elastic flux scales 
with $Z^2$ and $1/x$, dominating at small $x$ and for heavy nuclei. 
At large $x$, however, the inelastic component, driven by the proton 
and neutron photon PDFs, prevails. The $x$ value where both contributions become comparable depends on the hard scale $\mu$. These properties suggest that, in ultraperipheral collisions of heavy ions, photon-induced processes at large $x$ receive sizable semi-elastic and inelastic contributions.

The total photon inelastic flux in the nucleus is calculated as: 
%--------------------------------------------------------------------
\begin{equation}
x f_{\gamma/A}^{ine}(x) =  Z x f_{\gamma/p}(x) + N x f_{\gamma/n}(x) \; .
\end{equation}
%--------------------------------------------------------------------
In our evaluation we do not include any shadowing effect.

%----------------------------------------------------------------------
\begin{figure}[!h]
\begin{center}
     \includegraphics[scale=0.3]{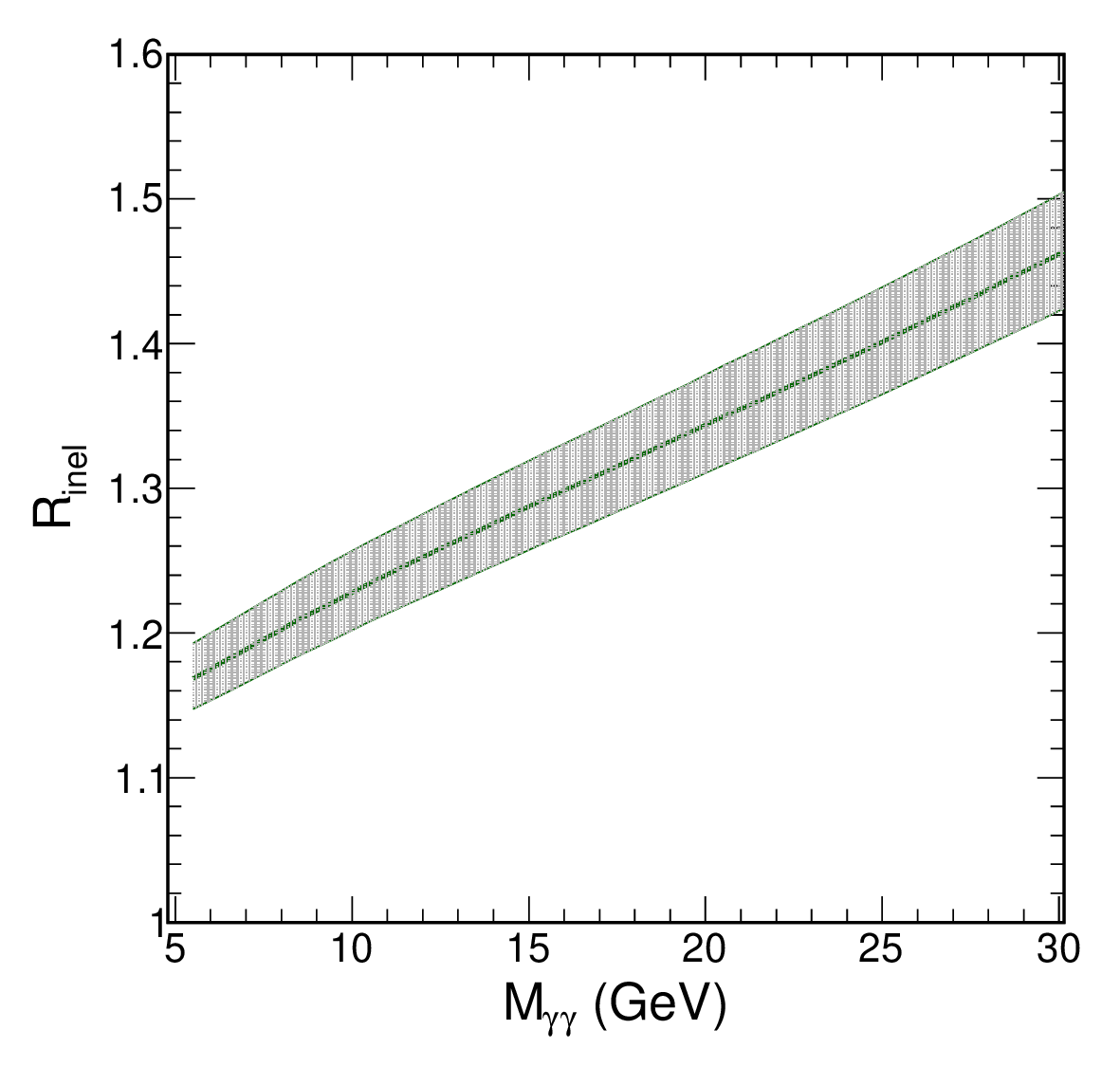}
     \includegraphics[scale=0.3]{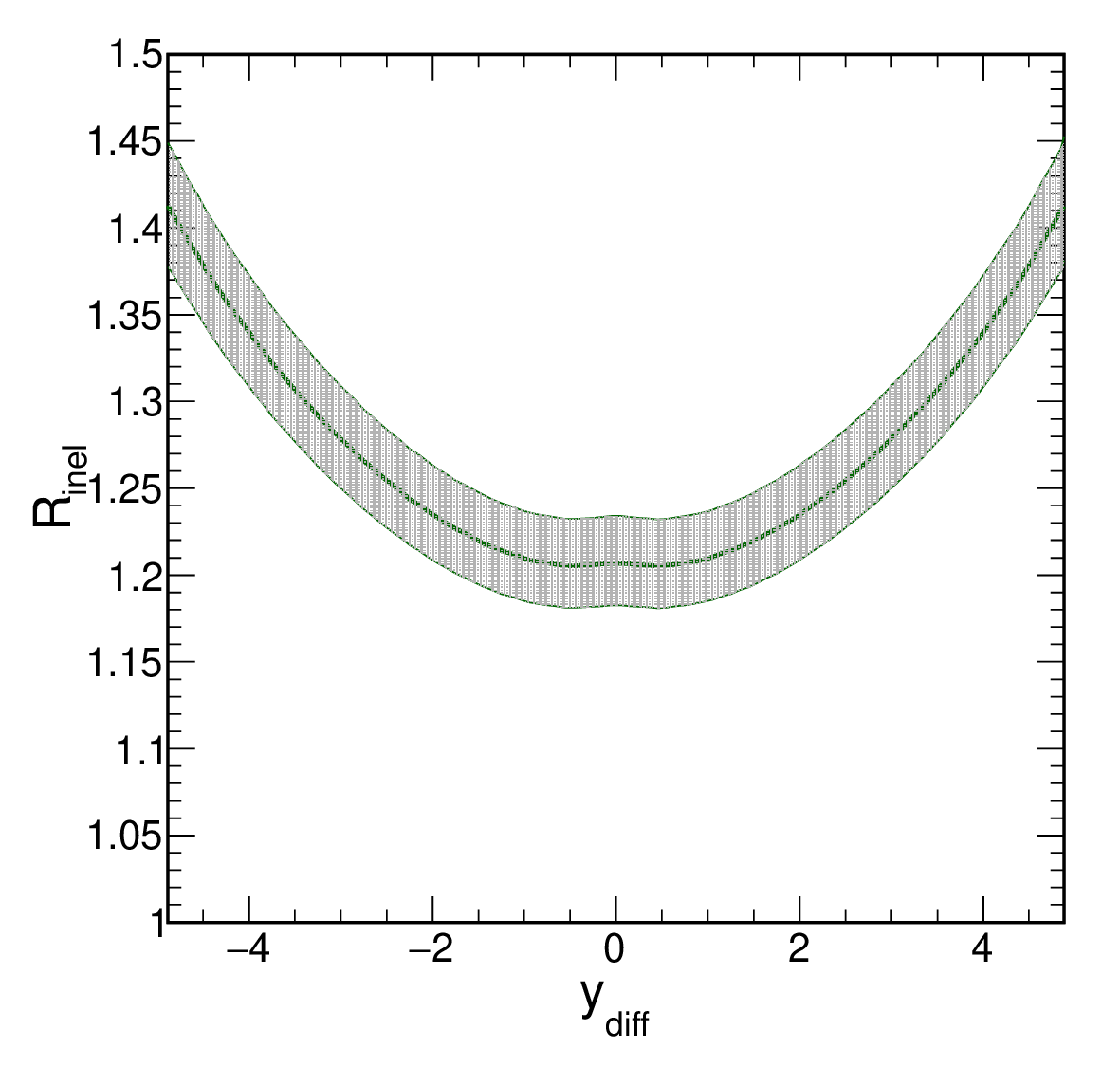}
\end{center}
\caption{The ratio of the modified cross section 
to the standard coherent-coherent contribution.
$R_{inel} = \frac{d \sigma^{lbl}(elastic+semi-elastic+inelastic)}
                 {d \sigma^{lbl}(elastic)}$,
$\mu^2 \in (M_{\gamma \gamma}/2, 2 M_{\gamma \gamma})$.}
\label{fig:ratios}
\end{figure}
%---------------------------------------------------------------------

How big are inelastic contributions compared to the standard ones
is shown in Fig. \ref{fig:ratios} as a function of 
$M_{\gamma \gamma}$ (left panel)
and $y_{diff}$ (right panel).
The results demonstrate a clear dependence on both variables and 
indicate sizable semi-elastic and inelastic contributions in 
the ATLAS kinematic range. An uncertainty band is also included. 
It was obtained by varying the factorization scale between 
$M^2_{\gamma\gamma/2}$ and $2M^2_{\gamma\gamma}$.
We do not discuss here how one could try to eliminate the inelastic
contributions. This will be discussed elsewhere.

%----------------------------------------------------------------
\begin{figure}[!h]
\begin{center}
     \includegraphics[scale=0.3]{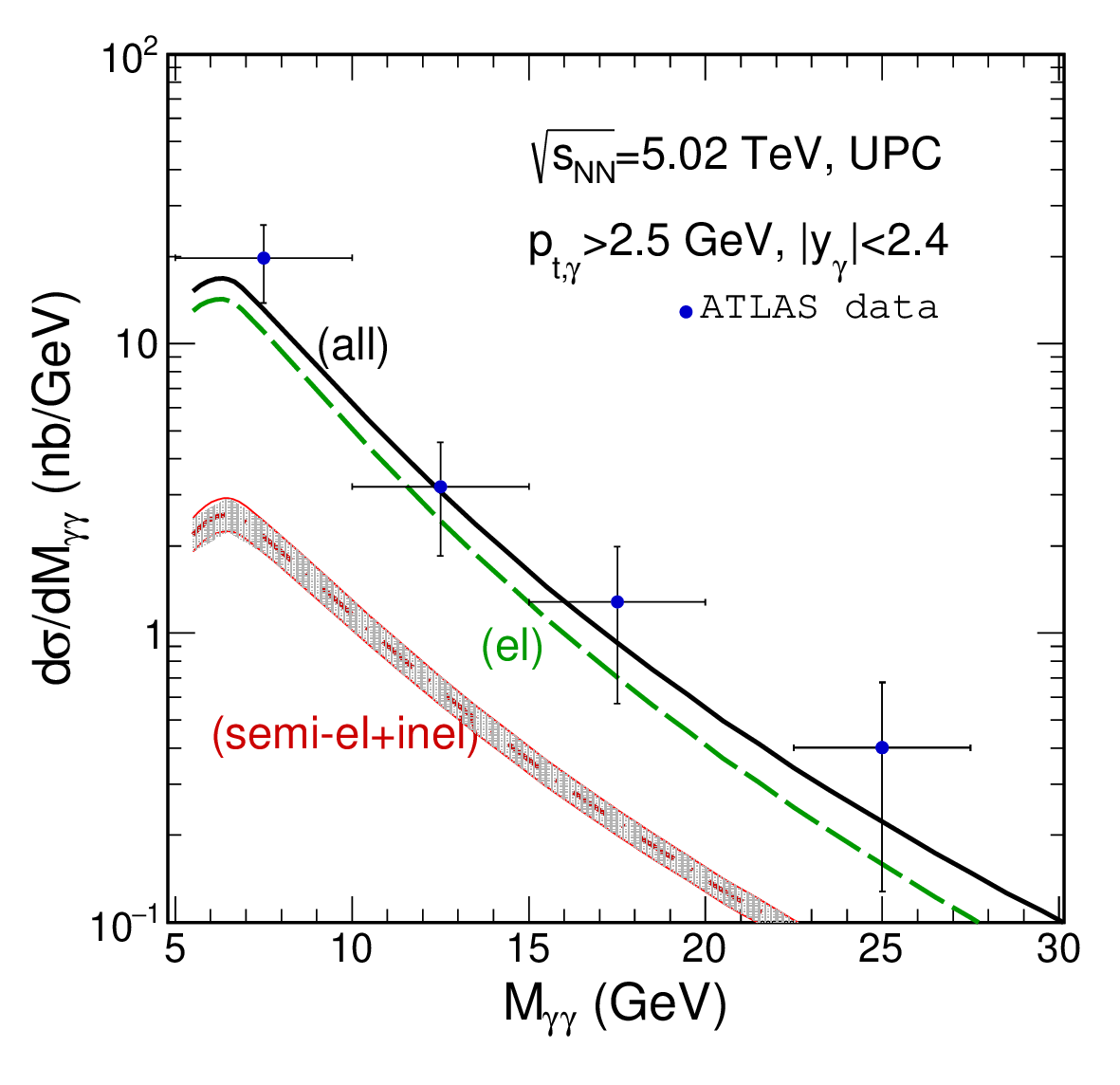}
\end{center}
\caption{Diphoton invariant mass distribution. Here we have naively 
added elastic and inelastic contributions.}
\label{fig:summary}
\end{figure}
%----------------------------------------------------------------

Fig. \ref{fig:summary} shows both elastic and inelastic contributions
together with ATLAS experimental data \cite{ATLAS:2020hii}.
Shown separately are the elastic (green dashed) and the semi-elastic 
plus inelastic (red dashed-dotted) contributions. As anticipated 
from the ratio analysis, the latter amount to about $20–40\%$ of 
the fully elastic part at small $M_{\gamma\gamma}$. 
The total result (black solid) provides a slightly improved description of the data.
The uncertainties illustrated in Fig. \ref{fig:summary} for 
inelastic contributions are due to variation of factorization scales
in the interval $\mu^2 \in (\mu^2/2, 2 \mu^2)$.

%--------------------------
\section{Conclusions}
%--------------------------

Several mechanisms of $\gamma \gamma \to \gamma \gamma$ 
scattering have been mentioned and discussed shortly
in the presentation at the conference.
We have discussed shortly double photon fluctuation into vector mesons
and Regge-type of interaction (VDM-Regge) as well as its interference 
with boxes. Negative interference was found.

One observes a missing strength with respect to the ATLAS (CMS) data.
What is possible reason for this?
Next-to-leading order effects were discussed in the talk
by Shao during the conference (see also \cite{AH:2023ewe,AH:2023kor}).
The role of fully charm tetraquark contribution was also discussed 
in the literature.
The mechanisms seem to be insufficient to explain the disagreement
of theoretical results with experimental data.

In our recent paper \cite{GKS2025} we discussed new inelastic 
contributions. 
In these mechanisms at least one photon couples to the nucleon 
(proton or neutron).
In our calculation we used CTEQ18QED collinear photon distributions
in proton and neutron.
We have shown that the inelastc contributions constitute 20-40 \%
of fully elastic contribution. The size depends on kinematics
($M_{\gamma \gamma}, y_{diff}$).
We also discussed uncertainties due to choice of 
the factorization scale.

In future estimation of shadowing effects would be very useful.
Modelling final state (associated emission of neutrons
and protons) excited by truely inelastic processes
is another open issue.
Can one distinguish final state excited by extra photon exchange 
from that due to one-step inelastic process.
Can the inelastic $\gamma \gamma \to \gamma \gamma$ 
processes be measured?
This requires extra studies, including modelling of gap survival 
factor.
In collinear approach to inelastic processes (as here) 
$p_t(\gamma \gamma) =$ 0,
but including $k_t$'s of photons $p_t(\gamma \gamma) \ne$ 0.
To understand the issue better one could vary the cut on 
$p_t(\gamma \gamma)$ or accoplanarity.
Can the measurement of neutrons in ZDC help 
in identifying the inelastic processes? 
This is e.g. the case for $J/\psi$ production \cite{SZGK}.
Very recently we studied production of neutrons associated
with $\gamma \gamma \to \gamma \gamma$ scattering \cite{KGJSz}.
A discrepancy from our predictions may signal presence
of the inelastic contributions dicussed here.

%---------------------------------
\section*{Acknowledgments}
%---------------------------------

We are indebted to Victor Goncalves for the collaboration
on the topic presented here.

%--------------------------
\section*{References}
%--------------------------

\end{document}